# Assessment of GaPSb/Si tandem material association properties for photoelectrochemical cells

Lipin Chen[1], Mahdi Alqahtani [2,4], Christophe Levallois[1], Antoine Létoublon[1], Julie Stervinou[1], Rozenn Piron[1], Soline Boyer-Richard[1], Jean-Marc Jancu[1], Tony Rohel[1], Rozenn Bernard[1], Yoan Léger[1], Nicolas Bertru[1], Jiang Wu[2], Ivan P. Parkin[3], Charles Cornet[1*]

[1]Univ Rennes, INSA Rennes, CNRS, Institut FOTON – UMR 6082, F-35000Rennes, France
[2]Department of Electronic and Electrical Engineering, University College London, London WC1E 7JE, United Kingdom
[3]Department of Chemistry, University College London, London WC1H 0AJ, United Kingdom
[4]King Abdulaziz City for Science and Technology
⋆E-mail: Charles.Cornet@insa-rennes.fr

Here, the structural, electronic and optical properties of the $GaP_{1-x}Sb_x$/Si tandem materials association are determined in view of its use for solar water splitting applications. The GaPSb crystalline layer is grown on Si by Molecular Beam Epitaxy with different Sb contents. The bandgap value and bandgap type of GaPSb alloy are determined on the whole Sb range, by combining experimental absorption measurements with tight binding (TB) theoretical calculations. The indirect (X-band) to direct (Γ-band) cross-over is found to occur at 30% Sb content. Especially, at a Sb content of 32%, the $GaP_{1-x}Sb_x$ alloy reaches the desired 1.7eV direct bandgap, enabling efficient sunlight absorption, that can be ideally combined with the Si 1.1 eV bandgap. Moreover, the band alignment of $GaP_{1-x}Sb_x$ alloys and Si with respect to water redox potential levels has been analyzed, which shows the GaPSb/Si association is an interesting combination both for the hydrogen evolution and oxygen evolution reactions. These results open new routes for the development of III-V/Si low-cost high-efficiency photoelectrochemical cells.

1. Introduction

The conversion of solar energy into green hydrogen fuel is one significant milestone on the road to a sustainable energy future [1,2]. Especially under the current background of global energy and environmental crisis, the development of the photoelectrochemical (PEC) water splitting technology, where the sunlight turns the liquid water into gaseous storable hydrogen that can be reused on demand for heat or electricity production, has driven many researches in the past years [3].

However, numbers of challenges remain for the development of this technology, in terms of efficiency, profitability and sustainability. The heart of the PEC conversion process is the choice of an appropriate photoelectrode material with both good bandgap and good band alignment to harvest the largest portion of the solar spectrum and provide sufficient voltage to accomplish the water splitting reactions [4]. Due to the broadness of the solar spectrum, from infrared to ultra-violet light, the needs for combining monolithically different materials with different bandgaps and thus absorbing different wavelengths, is considered today as the main pathway to reach high solar-to-hydrogen (STH) conversion efficiency [5,6]. In the tandem association of two materials, the optimum combination of bandgaps has been widely discussed for different device configurations [7]. Especially, the combination of a 1.7 eV top absorber bandgap with a 1.1 eV bottom absorber is recognized as one of the best tandem materials configuration and gives rise to a theoretical maximum STH efficiency $\eta_{STH}$ larger than 26% [8]. Different device demonstrations were proposed with a III-V/III-V materials design, mostly based on the Ga(In)As/GaInP association, reaching $\eta_{STH}$ larger than 15% [8-11]. In these works, high quality materials are achieved, thanks to the lattice-matching of theses alloys on the expensive GaAs substrate. A way to reduce the cost of tandem materials association is to monolithically integrate a III-V top absorber (1.7 eV bandgap) on the silicon substrate (because of its approximate 1.1eV band gap, earth abundance, low cost and prevalence in the electronics and PV industries) [7]. But by now there are only very few reports on Si-based tandem systems for water splitting, mainly due to the difficulty in growing high quality III-V epilayers on Si. The recent progress in the understanding of III-V/Si epitaxial processes and devices developments gives new hopes for the development of high efficiency III-V/Si PEC devices on the low-cost Si substrate [12,13]. Following this approach, pioneering works were performed on the development of bipolar configured 1.6eV AlGaAs on 1.1eV Si tandem association. Water splitting at a 18.3% conversion efficiency was reported [14]. Very recently, a single InGaN absorber photoanode monolithically integrated on silicon (where the Si (111) substrate was used as a back contact) was proposed with an applied-bias photo-to-current efficiency of 4.1% [15]. GaP-based materials were also proposed as an ideal tandem association with Si for water splitting [16]. The optimum design of GaP/Si tunnel junctions was even considered [17]. $TiO_2$-, $CoO_x$- or Ni-based passivation strategies were found to be efficient for photoelectrode stability [18,19]. But this approach still suffers from the indirect and large bandgap (2.26 eV) nature of GaP. Therefore, alloying GaP with other group III-V atoms is needed to achieve a direct bandgap at 1.7 eV. Doscher H. et. al theoretically proposed GaPN/Si and GaPNAs/Si tandem lattice-matched materials for water splitting based on the experiments of GaP/Si photoelectrochemistry and GaPN/Si epitaxial growth [20]. But these N-including lattice-matched materials face the issue of excitons localization effects, which hamper easily charge transport in the developed devices. On the other hand, metamorphic III-V integration on Si is known to generate dislocations that may propagate in the volume and are detrimental for solar devices. Recently, the synthesis of GaPSb was proposed to reduce the bandgap of GaP, with promising properties for PEC operation [21]. The epitaxial growth of Sb-based III-V compounds on Si substrates is particularly interesting in this regard, as it allows relaxation of the crystal stress through a near-perfect misfit dislocation network localized at the III-V/Si interface [12] leading to efficient and stable photonic device demonstrations [22]. Recent work demonstrated efficient operation of a GaPSb/Si photoanode for water splitting, but a complete assessment of the alloy with different compositions for water splitting was not yet given [23].

In this study, we evaluate the potential of $GaP_{1-x}Sb_x$/Si tandem materials association for the development of efficient III-V/Si photoelectrodes. $GaP_{1-x}Sb_x$ alloys with different compositions were directly grown on Si substrate by molecular beam epitaxy (MBE). The bandgaps and band alignments of the $GaP_{1-x}Sb_x$ alloys were carefully and comprehensively studied over the whole Sb range, by combining the experimental data with tight binding (TB) theoretical calculations. We finally discuss the promises offered by such alloy for its use in photoelectrodes.

2. Material and device design for solar water splitting

Three $GaP_{1-x}Sb_x$/Si samples (named as GaPSb-1, GaPSb-2 and GaPSb-3, with increasing Sb amounts) were grown by Molecular Beam Epitaxy (MBE) on HF-chemically prepared n-doped ($10^{17}cm^{-3}$) Si(001) substrates, with a 6° miscut toward the [110] direction[24]. The substrates were heated at 800°C for 10 minutes to remove hydrogen at the surface. 1μm-thick $GaP_{1-x}Sb_x$ layers were then grown at 500°C in a conventional continuous MBE growth mode, and at a growth rate of 0.24 ML/s, with a Beam Equivalent Pressure V/III ratio of 5.

The schematic diagrams of the proposed $GaP_{1-x}Sb_x$/Si tandem device for water splitting and the light absorption of the $GaP_{1-x}Sb_x$/Si tandem system with around 1.7/1.1eV bandgap combination are shown in Fig.1a and Fig.1b. The sun light first enters the top cell (1.7eV-bandgap targeted) $GaP_{1-x}Sb_x$ layer, in which high-energy photons are absorbed and low-energy photons are transmitted and harvested by the Si substrate or bottom cell, leading to an overall very large light absorption (as shown in Fig.1b). Then the photo-induced charges (electrons and holes) are generated in both layers, and, depending on the doping, one kind of charges flows toward the illuminated surface (the top surface of GaPSb layer) to generate $H_2$ or $O_2$ and the other one flow toward the back contact which is connected to the Si substrate (as shown in Fig.1a) and is further extracted to feed the counter-electrode.

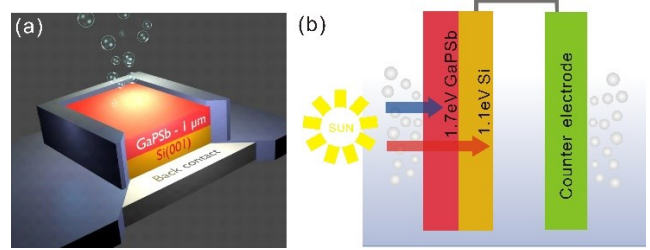

Fig. 1. (a) Schematic of the proposed $GaP_{1-x}Sb_x$/Si tandem device for PEC water splitting. (b) Sunlight absorption illustration of $GaP_{1-x}Sb_x$/Si tandem system with around 1.7/1.1eV bandgap combination for water splitting which can realize high light absorption.

3. Material characterizations and theoretical calculations

The detailed description of the methods for material characterizations and theoretical calculations are given in the supplemental materials.

Figure 2 shows the X-ray diffraction (XRD) patterns for the three samples. The miscut of around 6° is observed toward the [110] direction for the three samples based on the positions of Si Bragg peaks, which are in agreement with the substrate specifications. The ω/2θ scans exhibit well-defined $GaP_{1-x}Sb_x$ Bragg peaks for the three samples. Reciprocal space maps (RSM) carried out on either (004) (the insets of Fig.2a, Fig.2b, Fig.2c) or (115) (Fig.S1 in supplemental materials) reflections show a full plastic relaxation of the $GaP_{1-x}Sb_x$ layers for the three samples. $GaP_{1-x}Sb_x$ lattice parameters were extracted from both RSM and ω/2θ scans, leading to very similar values for each sample, confirming the full plastic relaxation rates and giving mean lattice parameters of 0.5665 nm, 0.5835 nm, 0.6093 nm, for the sample GaPSb-1, GaPSb-2, GaPSb-3, respectively. The Sb contents of 0.33, 0.60, and >0.99 are then inferred [25]. Sample GaPSb-3 is almost pure GaSb. These RSM images also exhibit an important Bragg peak broadening due to a relatively large crystal defect density, in low Sb content samples. Furthermore, the XRD analysis does not give any evidence of a phase separation that could occur between GaP and GaSb in the $GaP_{1-x}Sb_x$ alloys in theses growth conditions.

Figure 3 shows the scanning electron microscopy (SEM) images of the three $GaP_{0.67}Sb_{0.33}$/Si, $GaP_{0.40}Sb_{0.60}$/Si, GaSb/Si samples on plane-view (Fig.3 a, c, e) and cross-section view (Fig.3 b, d, f). From these images, it can be observed that the samples $GaP_{0.40}Sb_{0.60}$/Si and GaSb/Si exhibit relatively smooth surfaces as compared with the other sample, the $GaP_{0.67}Sb_{0.33}$/Si. The roughness observed is attributed to both emergence of some crystal defects such as residual dislocations and residual stress in the sample. The corresponding atomic force microscopy (AFM) images are given in the supplemental materials and the RMS (root-mean-square) roughnesses of the surfaces were found to be 22.80nm for $GaP_{0.67}Sb_{0.33}$, 9.32nm for $GaP_{0.40}Sb_{0.60}$ and 7.91nm for GaSb. It suggests an improvement of the crystal quality with increasing Sb content. This is also confirmed by a very significant Bragg peak sharpening observed on RSM (Fig.2). Nevertheless, the crystal quality of the GaPSb alloy grown on Si has been significantly improved compared with the one obtained in previous work [21].

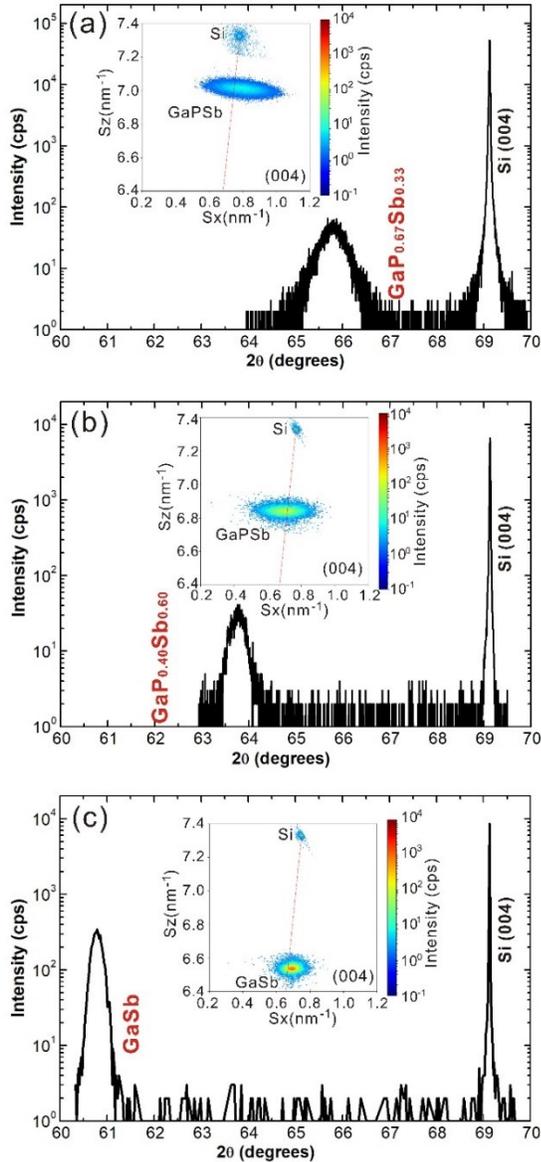

Fig. 2. X-Ray Diffraction patterns of three MBE-grown $GaP_{1-x}Sb_x$/Si samples with different Sb contents: GaPSb-1 (a), GaPSb-2 (b) and GaPSb-3 (c). The insets show the reciprocal space maps around (004) for the three samples, correspondingly (Sx and Sz are the projected coordinates in the right handed Cartesian, with z axis parallel to the surface normal) [25].

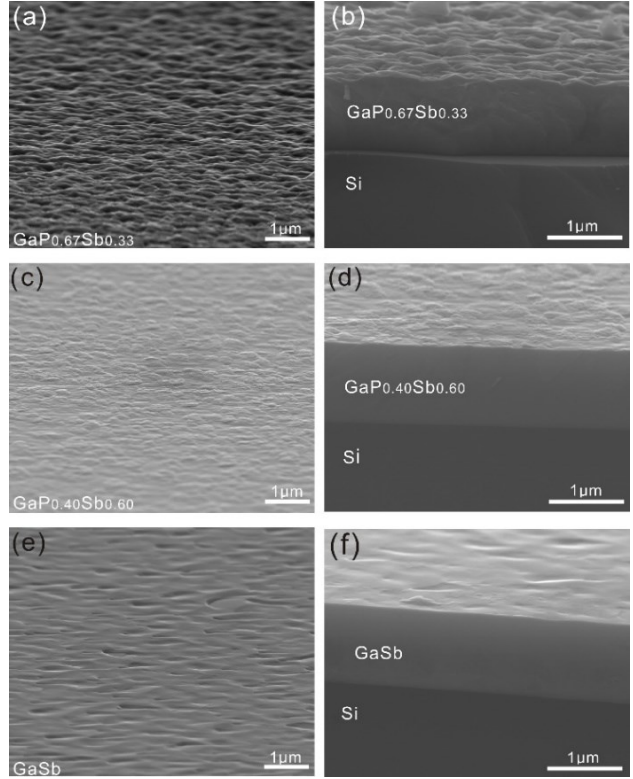

Fig. 3. Scanning electron microscopy (SEM) images of the $GaP_{0.67}Sb_{0.33}$, $GaP_{0.40}Sb_{0.60}$, GaSb three samples on plane view (a),(c),(e) and cross-section view(b),(d),(f).

Ellipsometry measurements were made for GaPSb bandgap determination, independently of the Si substrate. Chemical mechanical polishing (CMP) was performed to obtain smoother surfaces and avoid distortion on ellipsometry measurements. The SEM plane-view images of the three CMP-polished samples are shown in Fig.S2 (supplemental materials), which shows that the surface roughnesses of the three samples are much lower than the as-grown samples. The surface RMS roughnesses of the three CMP samples extracted from AFM measurements are 0.33nm ($GaP_{0.67}Sb_{0.33}$), 0.28nm ($GaP_{0.40}Sb_{0.60}$) and 0.56nm (GaSb), indicating the surfaces of the three samples become very smooth after CMP processes. Then the optical constants of the three $GaP_{0.67}Sb_{0.33}$/Si, $GaP_{0.40}Sb_{0.60}$/Si, and GaSb/Si samples were measured by variable angle spectroscopic ellipsometry (VASE) at room temperature in the 0.58–5 eV photon energy region. The angles of incidence were set to 60° and 70°. A Tauc-Lorentz model with two oscillators was used to fit the ellipsometry data of the three samples (Fig.S5 in the supplemental materials). From this model, the refractive index (n), extinction coefficient (k) and absorption spectrum of the GaPSb layers of the three samples were extracted independently of the Si substrate, respectively, as shown in Fig.S6 (supplemental materials) and Fig.4 (the red curves). Besides, in order to further integrate and verify the experimental data, ellipsometry measurement was also performed on a GaP/Si sample and the corresponding optical constants extracted based on the Tauc-Lorentz model are shown in Fig.S7 (supplemental materials). The deduced absorption curves of the GaP and GaSb based on ellipsometry measurements show good agreements with both the experimental and theoretical data presented in ref. [26] (see Fig.S8 in the supplemental materials).

Based on Tauc plot method, the band gap ranges of the four samples were obtained: Eg=2.25±0.04 eV for GaP; Eg=1.70±0.06 eV for $GaP_{0.67}Sb_{0.33}$; Eg=1.04±0.08 eV for $GaP_{0.40}Sb_{0.60}$, and Eg=0.68±0.09 eV for GaSb (the details see the supplemental materials), which are consistent with bandgaps reported for the metamorphic growth of GaPSb on InP substrate [27].

The band structures of the unstrained $GaP_{1-x}Sb_x$ alloys in the whole Sb compositional range have then been calculated by tight-binding calculation, using an extended basis sp3d5s* tight binding Hamiltonian [28]. From the tight binding parameters of GaP and GaSb binary compounds [28], a virtual crystal approximation is performed to obtain the band structures of the different $GaP_{1-x}Sb_x$ alloys at 0K as a function of the Sb content. Then the energies were shifted to take into account the influence of the temperature, in order to get room temperature band structures [29]. The calculated band structures of GaP, $GaP_{0.67}Sb_{0.33}$, $GaP_{0.40}Sb_{0.60}$, and GaSb semiconductors are given in the Fig.S10]. The bandgap evolutions obtained in the Γ, L and X valleys of the $GaP_{1-x}Sb_x$ alloy for the whole Sb content range are presented in Fig.5, from which we can find both the band gap value and the band gap type (direct or indirect). For the GaP, $GaP_{0.67}Sb_{0.33}$, $GaP_{0.40}Sb_{0.60}$, and GaSb semiconductors, TB calculations are in good agreement with the experimental absorption measurements. The theoretical absorption curves of $GaP_{0.67}Sb_{0.33}$, $GaP_{0.40}Sb_{0.60}$, GaSb semiconductors (corresponding to the three MBE grown samples) determined with TB calculations are shown as black lines in Fig.4, which also shows good consistency with the experimental data. From this analysis, we deduce that the Sb content at which a 1.7 eV direct bandgap needed for tandem materials association obtained is 32%.

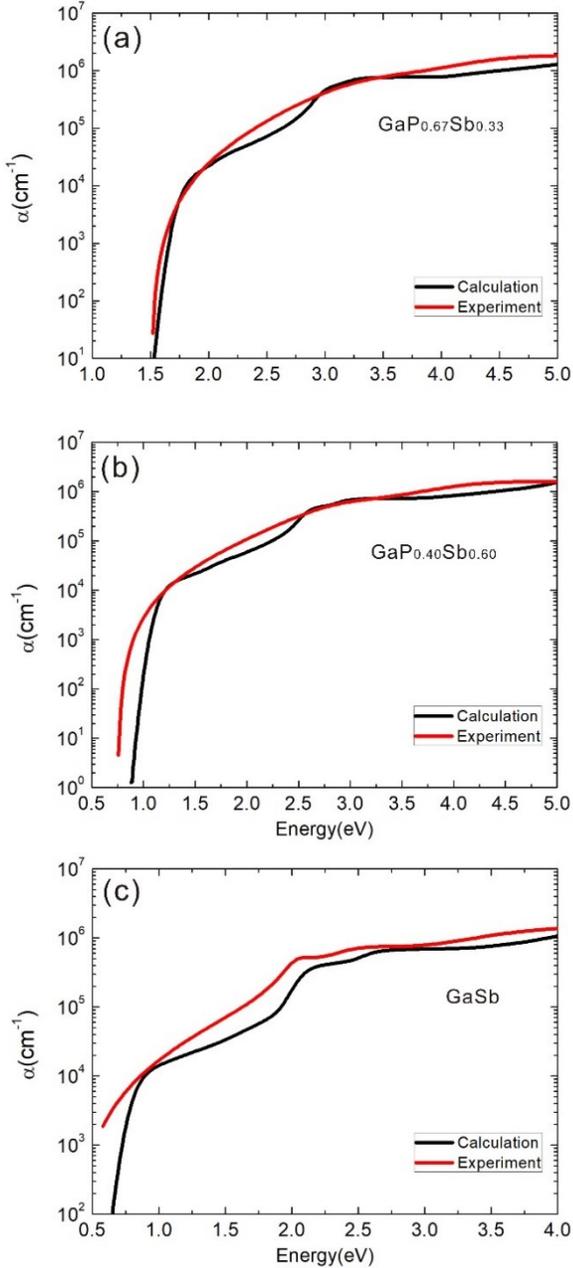

Fig. 4. Optical absorption spectra of the $GaP_{0.67}Sb_{0.33}$ (a), $GaP_{0.40}Sb_{0.60}$ (b), GaSb (c) semiconductors. The red lines were deduced from ellipsometry measurement and the black lines were obtained based on TB calculation.

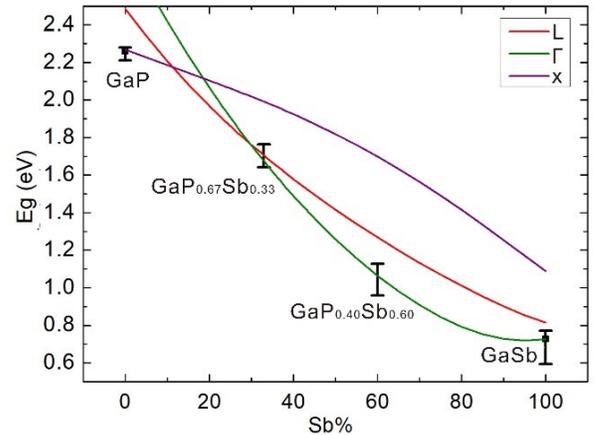

Fig. 5. Room temperature bandgaps of $GaP_{1-x}Sb_x$ alloys with different Sb contents. The green, red and purple solid lines are theoretical curves computed by tight-binding calculations, corresponding to Γ, L and X valleys, respectively. The separated error bar lines correspond to the bandgaps determined experimentally. The black dots correspond to the bandgaps of GaP and GaSb given in ref. [30,31].

For efficient PEC water splitting, another important aspect is to estimate the band edge location with respect to the redox potentials of water. The band alignment (no external electric field applied) of the $GaP_{1-x}Sb_x$/Si tandem architecture for water splitting is presented in Fig.6 as a function of the Sb content. The VBM (valence band maximum) energy (Ev) of the $GaP_{1-x}Sb_x$ alloy over the whole Sb content range was obtained based on the absolute bandlineup between VBM energy of GaP and GaSb [32], which follows the linear formula $Ev(GaP_{1-x}Sb_x)= xEv(GaSb)+ (1-x)Ev(GaP)$ [33]. In a first approximation, the evolution of the surface acidity with Sb content was neglected. The CBM (conduction band minimum) energy of the $GaP_{1-x}Sb_x$ alloy was obtained by using the results from TB calculations presented in Fig. 5. For this alloy, the minimum of the conduction band is located in the X-valley between 0% and 11%. It then moves to the L valley for Sb contents between 11% and 30%. It finally reaches a direct bandgap configuration (minimum of the CB in the Γ valley) beyond 30% of Sb incorporation. While the precise value predicted by the calculations for the X to L valleys crossover seems hard to confirm experimentally in this work, calculations clearly demonstrate that in any cases, an indirect to direct cross-over is expected for this alloy at around 30% Sb, contrary to previously calculated band structures using density functional theory [21]. The VBM energy of bulk GaP is very close to the oxidation potential of water, which has been verified by many reports [32,34-36]. While, in the different papers, the energy differences (between the water redox levels and GaP VBM energy) have a little difference. So, here we show the redox levels of water based on the VBM position of GaP [32] and the energy differences [32,34-36] with error bars to analyze potential water splitting reactions more accurately and comprehensively (Fig.6).

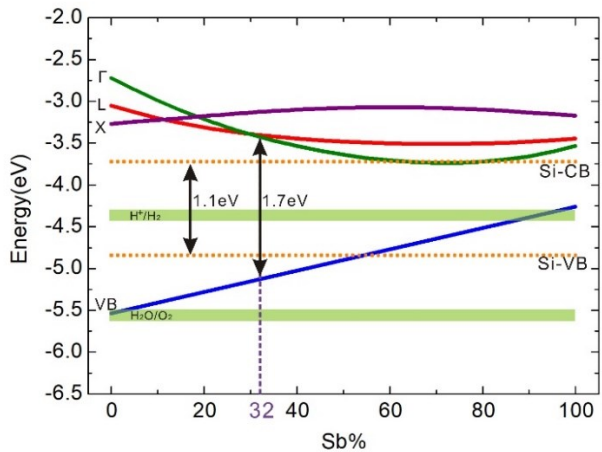

Fig. 6. Band alignment of Γ, L and X valleys of $GaP_{1-x}Sb_x$ and indirect bandgap of Si with respect to water oxidation and reduction potentials. The relative position of redox potentials of water are shown as green bars at pH=0.

From the Fig.5 and Fig.6, it can be observed first that, in order to benefit from a direct GaPSb bandgap larger than the one of the silicon (to absorb higher energy solar radiations), the Sb content should lie between 30% (indirect to direct cross-over) and 54% (bandgap equal to the Si one). In this Sb content range, it is also noticed that the bandlineup is of type I, promoting the charge carrier extraction in the silicon under zero bias conditions. Now looking at the bandlineups between GaPSb and the water redox potentials in the 30%-54% Sb content range, it is seen that GaPSb has a strong reduction ability due to its relatively higher CBM and it is around 0.9 eV higher than $H^+/H_2$ potential at 32% Sb content. On the other hand, the VBM is higher than the $O_2/H_2O$ potential in the 30%-54% Sb content range, and due to the linear increase of VBM with the Sb content, the GaPSb with low Sb content (around 30%) is thus more suitable for oxygen evolution reactions (about 0.43 and 0.46 eV higher than $O_2/H_2O$ potential at 30% and 32% Sb content). Overall, addition of Sb in GaP will therefore ease photocathode operation, although large currents densities were obtained with a GaPSb/Si photoanode, due to a strong direct bandgap absorption in this alloy [23]. The band lineups calculated in this work thus give a basis for further devices analysis, where the precise determination of Nernstian shifts and band bending will be needed, but it is beyond the scope of this article.

4. Conclusions

In summary, the structural, electronic and optical properties of the III-V/Si-based tandem materials association $GaP_{1-x}Sb_x$/Si for PEC water splitting was studied comprehensively for the whole range of Sb composition, and its potential for photoelectrochemical cell evaluated. The $GaP_{1-x}Sb_x$ alloys were directly grown on the Si substrate with different Sb contents. The bandgap values and bandgap types of GaPSb alloys were determined on the whole Sb range, by combining the experimental data with tight binding (TB) theoretical calculations. The indirect (X-band) to direct (Γ-band) cross-over was found to occur at 30% Sb content. Especially, at a Sb content of 32%, the $GaP_{1-x}Sb_x$ alloy reached the ideal 1.7eV direct bandgap, complementary to the Si 1.1eV one. Furthermore, the analysis of the band alignment of $GaP_{1-x}Sb_x$ alloys and Si with respect to water potential levels shows that the GaPSb/Si association is an interesting combination both for the hydrogen evolution and oxygen evolution reactions, suggesting the $GaP_{1-x}Sb_x$/Si tandem material holds great promise for high-efficiency solar water splitting on the low cost silicon substrate.

**Acknowledgement**

This research was supported by the French National Research Agency ANTIPODE Project (Grant No. 14-CE26-0014-01), Région Bretagne and King Abdulaziz City for Science and Technology (KACST), Riyadh, Saudi Arabia. Lipin Chen acknowledges the China Scholarship Council (CSC) for her Ph.D financial support (No. 2017-6254). Mahdi Alqahtani acknowledges the support and scholarship from King Abdulaziz City for Science and Technology, Riyadh, Saudi Arabia. The authors acknowledge RENATECH (French Network of Major Technology Centers) within Nanorennes for technological support.

## References

[1] A.J. Bard, M.A. Fox, Artificial photosynthesis: solar splitting of water to hydrogen and oxygen, Acc. Chem. Res. 28 (1995) 141–145.

[2] N.S. Lewis, D.G. Nocera, Powering the planet: Chemical challenges in solar energy utilization, Proc. Natl. Acad. Sci. 103 (2006) 15729–15735.

[3] D. Kang, T.W. Kim, S.R. Kubota, A.C. Cardiel, H.G. Cha, K.-S. Choi, Electrochemical Synthesis of Photoelectrodes and Catalysts for Use in Solar Water Splitting, Chem. Rev. 115 (2015) 12839–12887. doi:10.1021/acs.chemrev.5b00498.

[4] R. van de Krol, M. Grätzel, eds., Photoelectrochemical Hydrogen Production, Springer US, 2012. https://www.springer.com/gp/book/9781461413790 (accessed January 31, 2019).

[5] J. Rongé, T. Bosserez, D. Martel, C. Nervi, L. Boarino, F. Taulelle, G. Decher, S. Bordiga, J. A. Martens, Monolithic cells for solar fuels, Chem. Soc. Rev. 43 (2014) 7963–7981. doi:10.1039/C3CS60424A.

[6] D. Bae, B. Seger, P.C. Vesborg, O. Hansen, I. Chorkendorff, Strategies for stable water splitting via protected photoelectrodes, Chem. Soc. Rev. 46 (2017) 1933–1954.

[7] S. Hu, C. Xiang, S. Haussener, A.D. Berger, N.S. Lewis, An analysis of the optimal band gaps of light absorbers in integrated tandem photoelectrochemical water-splitting systems, Energy Environ. Sci. 6 (2013) 2984–2993. doi:10.1039/C3EE40453F.

[8] J.L. Young, M.A. Steiner, H. Döscher, R.M. France, J.A. Turner, T.G. Deutsch, Direct solar-to-hydrogen conversion via inverted metamorphic multi-junction semiconductor architectures, Nat. Energy. 2 (2017) 17028. doi:10.1038/nenergy.2017.28.

[9] O. Khaselev, J.A. Turner, A Monolithic Photovoltaic-Photoelectrochemical Device for Hydrogen Production via Water Splitting, Science. 280 (1998) 425–427. doi:10.1126/science.280.5362.425.

[10] M.M. May, H.-J. Lewerenz, D. Lackner, F. Dimroth, T. Hannappel, Efficient direct solar-to-hydrogen conversion by in situ interface transformation of a tandem structure, Nat. Commun. 6 (2015) 8286. doi:10.1038/ncomms9286.

[11] W.-H. Cheng, M.H. Richter, M.M. May, J. Ohlmann, D. Lackner, F. Dimroth, T. Hannappel, H.A. Atwater, H.-J. Lewerenz, Monolithic Photoelectrochemical Device for Direct Water Splitting with 19% Efficiency, ACS Energy Lett. 3 (2018) 1795–1800. doi:10.1021/acsenergylett.8b00920.

[12] I. Lucci, S. Charbonnier, L. Pedesseau, M. Vallet, L. Cerutti, J.-B. Rodriguez, E. Tournié, R. Bernard, A. Létoublon, N. Bertru, A. Le Corre, S. Rennesson, F. Semond, G. Patriarche, L. Largeau, P. Turban, A. Ponchet, C. Cornet, Universal description of III-V/Si epitaxial growth processes, Phys. Rev. Mater. 2 (2018) 060401. doi:10.1103/PhysRevMaterials.2.060401.

[13] S. Chen, W. Li, J. Wu, Q. Jiang, M. Tang, S. Shutts, S.N. Elliott, A. Sobiesierski, A.J. Seeds, I. Ross, P.M. Smowton, H. Liu, Electrically pumped continuous-wave III–V quantum dot lasers on silicon, Nat. Photonics. 10 (2016) 307–311. doi:10.1038/nphoton.2016.21.

[14] S. Licht, B. Wang, S. Mukerji, T. Soga, M. Umeno, H. Tributsch, Efficient solar water splitting, exemplified by RuO2-catalyzed AlGaAs/Si photoelectrolysis, J. Phys. Chem. B. 104 (2000) 8920–8924.

[15] P. Kumar, P. Devi, R. Jain, S.M. Shivaprasad, R.K. Sinha, G. Zhou, R. Nötzel, Quantum dot activated indium gallium nitride on silicon as photoanode for solar hydrogen generation, Commun. Chem. 2 (2019) 4. doi:10.1038/s42004-018-0105-0.

[16] I. Lucci, S. Charbonnier, M. Vallet, P. Turban, Y. Léger, T. Rohel, N. Bertru, A. Létoublon, J.-B. Rodriguez, L. Cerutti, E. Tournié, A. Ponchet, G. Patriarche, L. Pedesseau, C. Cornet, A Stress-Free and Textured GaP Template on Silicon for Solar Water Splitting, Adv. Funct. Mater. 28 (2018) 1801585. doi:10.1002/adfm.201801585.

[17] A. Rolland, L. Pedesseau, J. Even, S. Almosni, C. Robert, C. Cornet, J.M. Jancu, J. Benhlal, O. Durand, A.L. Corre, P. Rale, L. Lombez, J.-F. Guillemoles, E. Tea, S. Laribi, Design of a lattice-matched III–V–N/Si photovoltaic tandem cell monolithically integrated on silicon substrate, Opt. Quantum Electron. 46 (2014) 1397–1403. doi:10.1007/s11082-014-9909-z.

[18] M. Alqahtani, S. Ben-Jabar, M. Ebaid, S. Sathasivam, P. Jurczak, X. Xia, A. Alromaeh, C. Blackman, Y. Qin, B. Zhang, B.S. Ooi, H. Liu, I.P. Parkin, J. Wu, Gallium Phosphide photoanode coated with $TiO_2$ and $CoO_x$ for stable photoelectrochemical water oxidation, Opt. Express. 27 (2019) A364–A371. doi:10.1364/OE.27.00A364.

[19] S. Hu, M.R. Shaner, J.A. Beardslee, M. Lichterman, B.S. Brunschwig, N.S. Lewis, Amorphous $TiO_2$ coatings stabilize Si, GaAs, and GaP photoanodes for efficient water oxidation, Science. 344 (2014) 1005–1009.

[20] H. Döscher, O. Supplie, M.M. May, P. Sippel, C. Heine, A.G. Muñoz, R. Eichberger, H.-J. Lewerenz, T. Hannappel, Epitaxial III–V films and surfaces for photoelectrocatalysis, ChemPhysChem. 13 (2012) 2899–2909.

[21] A. Martinez-Garcia, H.B. Russell, W. Paxton, S. Ravipati, S. Calero-Barney, M. Menon, E. Richter, J. Young, T. Deutsch, M.K. Sunkara, Unassisted Water Splitting Using a $GaSb_xP_{(1-x)}$ Photoanode, Adv. Energy Mater. (2018) 1703247.

[22] H. Nguyen-Van, A.N. Baranov, Z. Loghmari, L. Cerutti, J.-B. Rodriguez, J. Tournet, G. Narcy, G. Boissier, G. Patriarche, M. Bahriz, E. Tournié, R. Teissier, Quantum cascade lasers grown on silicon, Sci. Rep. 8 (2018) 7206. doi:10.1038/s41598-018-24723-2.

[23] M. Alqahtani, S. Sathasivam, L. Chen, P. Jurczak, R. Piron, C. Levallois, A. Létoublon, Y. Léger, S. Boyer-Richard, N. Bertru, J.-M. Jancu, C. Cornet, J. Wu, I.P. Parkin, Photoelectrochemical water oxidation of $GaP_{1-x}Sb_x$ with a direct band gap of 1.65 eV for full spectrum solar energy harvesting, Sustain. Energy Fuels. 3 (2019) 1720–1729. doi:10.1039/C9SE00113A.

[24] T. Quinci, J. Kuyyalil, T.N. Thanh, Y.P. Wang, S. Almosni, A. Létoublon, T. Rohel, K. Tavernier, N. Chevalier, O. Dehaese, Defects limitation in epitaxial GaP on bistepped Si surface using UHVCVD–MBE growth cluster, J. Cryst. Growth. 380 (2013) 157–162.

[25] T. Nguyen Thanh, C. Robert, W. Guo, A. Létoublon, C. Cornet, G. Elias, A. Ponchet, T. Rohel, N. Bertru, A. Balocchi, Structural and optical analyses of GaP/Si and (GaAsPN/GaPN)/GaP/Si nanolayers for integrated photonics on silicon, J. Appl. Phys. 112 (2012) 053521.

[26] S. Adachi, Optical dispersion relations for GaP, GaAs, GaSb, InP, InAs, InSb, $Al_xGa_{1-x}As$, and $In_{1-x}Ga_xAs_yP_{1-y}$, J. Appl. Phys. 66 (1989) 6030–6040. doi:10.1063/1.343580.

[27] S. Loualiche, A. Le Corre, S. Salaun, J. Caulet, B. Lambert, M. Gauneau, D. Lecrosnier, B. Deveaud, GaPSb: A new ternary material for Schottky diode fabrication on InP, Appl. Phys. Lett. 59 (1991) 423–424.

[28] J.-M. Jancu, R. Scholz, F. Beltram, F. Bassani, Empirical spds* tight-binding calculation for cubic semiconductors: General method and material parameters, Phys. Rev. B. 57 (1998) 6493.

[29] Y.P. Varshni, Temperature dependence of the energy gap in semiconductors, Physica. 34 (1967) 149–154.

[30] M.B. Panish, H.C. Casey, Temperature Dependence of the Energy Gap in GaAs and GaP, J. Appl. Phys. 40 (1969) 163–167. doi:10.1063/1.1657024.
[31] M. Wu, C. Chen, Photoluminescence of high- quality GaSb grown from Ga- and Sb- rich solutions by liquid- phase epitaxy, J. Appl. Phys. 72 (1992) 4275–4280. doi:10.1063/1.352216.
[32] C.G. Van de Walle, J. Neugebauer, Universal alignment of hydrogen levels in semiconductors, insulators and solutions, Nature. 423 (2003) 626.
[33] B.M. Borg, L.-E. Wernersson, Synthesis and properties of antimonide nanowires, Nanotechnology. 24 (2013) 202001.
[34] D. Jing, L. Guo, L. Zhao, X. Zhang, H. Liu, M. Li, S. Shen, G. Liu, X. Hu, X. Zhang, Efficient solar hydrogen production by photocatalytic water splitting: from fundamental study to pilot demonstration, Int. J. Hydrog. Energy. 35 (2010) 7087–7097.
[35] S. Chu, W. Li, Y. Yan, T. Hamann, I. Shih, D. Wang, Z. Mi, Roadmap on solar water splitting: current status and future prospects, Nano Futur. 1 (2017) 022001.
[36] A. Kudo, Y. Miseki, Heterogeneous photocatalyst materials for water splitting, Chem. Soc. Rev. 38 (2009) 253–278.

# Assessment of GaPSb/Si tandem material association properties for photoelectrochemical cells

Lipin Chen[1], Mahdi Alqahtani [2,4], Christophe Levallois[1], Antoine Létoublon[1], Julie Stervinou[1], Rozenn Piron[1], Soline Boyer-Richard[1], Jean-Marc Jancu[1], Tony Rohel[1], Rozenn Bernard[1], Yoan Léger[1], Nicolas Bertru[1], Jiang Wu[2], Ivan P. Parkin[3], Charles Cornet[1*]

[1]Univ Rennes, INSA Rennes, CNRS, Institut FOTON – UMR 6082, F-35000Rennes, France
[2]Department of Electronic and Electrical Engineering, University College London, London WC1E 7JE, United Kingdom
[3]Department of Chemistry, University College London, London WC1H 0AJ, United Kingdom
[4]King Abdulaziz City for Science and Technology
⋆E-mail: Charles.Cornet@insa-rennes.fr

**X-Ray Diffraction (XRD)**

The synthesized $GaP_{1-x}Sb_x$/Si samples were characterized by X-ray diffraction (XRD) on a 4 circles Brucker D8 Diffractometer. A Bartels asymmetric Ge (220) monochromator was used for both line scan and reciprocal space maps (RSM). The detection is ensured by a Lynxeye$^{TM}$, 1 dimensional position sensitive detector (PSD) allowing a collection angle of 2.6° over 2θ.

The Reciprocal space maps were carried out on either (004) (Fig.1 insets) and (115) (Fig.S1) reflections, which show a full plastic relaxation of the $GaP_{1-x}Sb_x$ layer for the three samples.

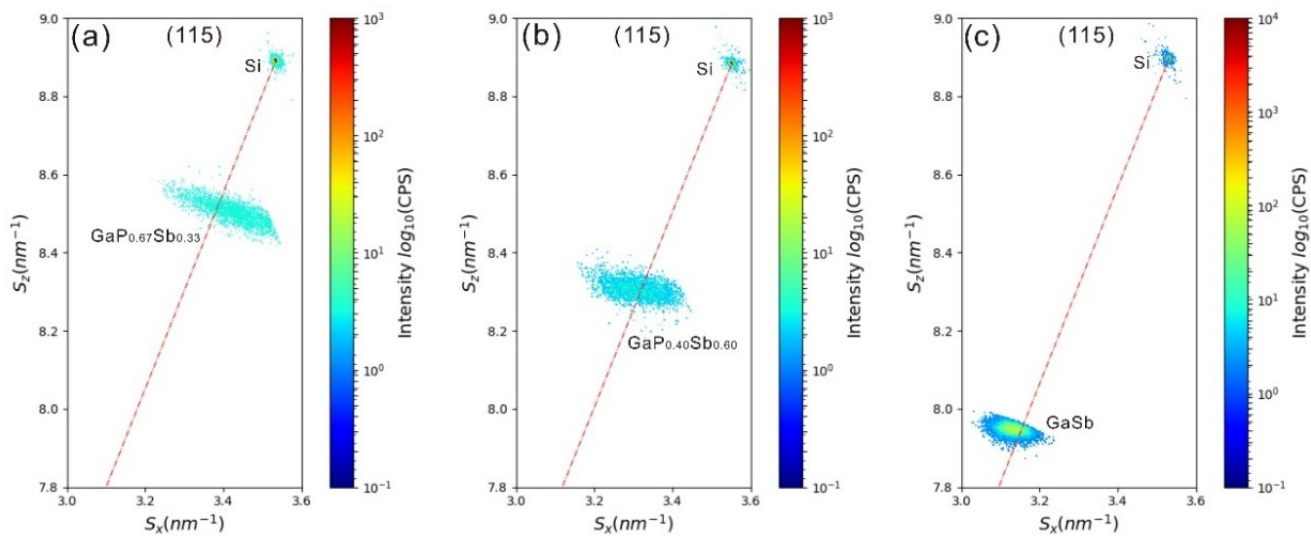

Fig. S1. X-Ray diffraction reciprocal space mappings around the (115) Bragg positions for the $GaP_{0.67}Sb_{0.33}$ (a), $GaP_{0.40}Sb_{0.60}$ (b), GaSb (c) samples. The red line represents the full relaxation line.

**Chemical Mechanical Polishing (CMP)**

The samples were polished by chemical mechanical polishing method with 1% $H_3PO_4$ etching solution at a rate 1 round/s for 30mins.

**Scanning Electron Microscopy (SEM)**

Scanning electron microscopy images were obtained by using a JEOL JSM-7100 scanning electron microscope.
Fig.S2 shows the plane-view SEM images of the $GaP_{0.67}Sb_{0.33}$/Si, $GaP_{0.40}Sb_{0.60}$/Si, GaSb/Si samples, from which we can find the surface quality of the three samples are much improved.

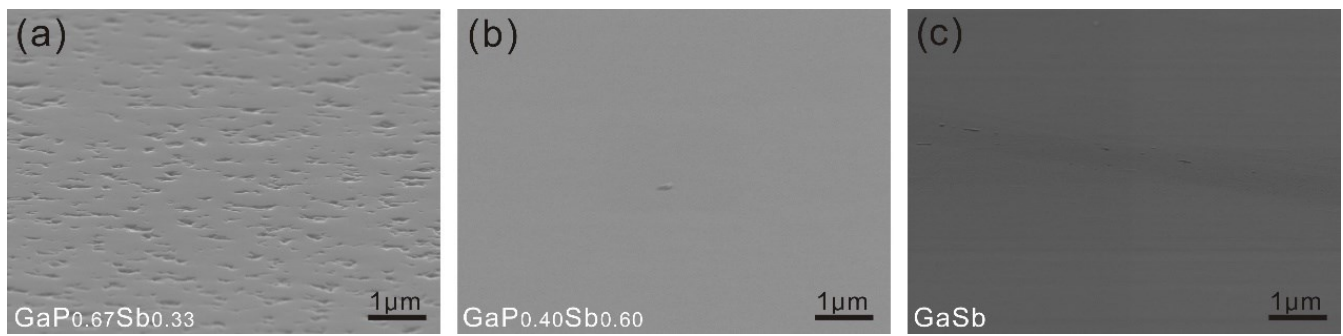

Fig. S2. Scanning electron microscopy (SEM) plane-view images of the GaP$_{0.67}$Sb$_{0.33}$ (a), GaP$_{0.40}$Sb$_{0.60}$ (b), GaSb (c) samples after CMP.

**Atomic Force Microscopy (AFM)**

Atomic force microscopy (AFM) measurements were performed based on a Veeco Innova AFM microscope with a high-resolution scanning probe. Tapping mode was used with the cantilever tuned around 293KHz.

The atomic force microscopy (AFM) measurements were made to study the surface roughness of the three samples before and after CMP quantitatively. Figure S3 shows the AFM images of the three as-grown samples. The RMS (root-mean-square) roughness of the surfaces were calculated at 22.80 nm for GaP$_{0.67}$Sb$_{0.33}$, 9.32nm for GaP$_{0.40}$Sb$_{0.60}$ and 7.91nm for GaSb. Figure S4 shows the AFM images of the three samples after CMP and the surface RMS roughness of the three CMP samples are 0.33nm (GaP$_{0.67}$Sb$_{0.33}$), 0.28nm (GaP$_{0.40}$Sb$_{0.60}$) and 0.56nm (GaSb), indicating the surface of the three samples become very smooth after CMP processes.

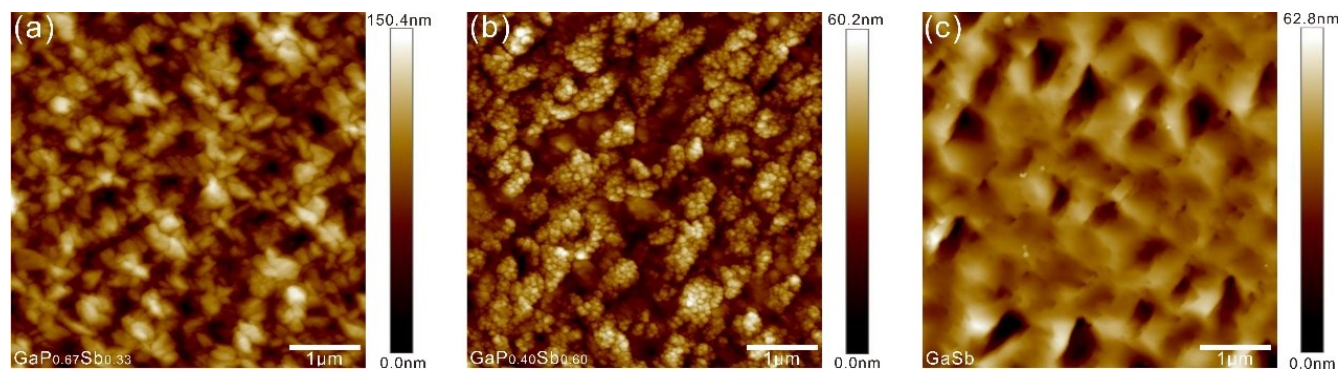

Fig. S3. Atomic force microscopy (AFM) images of the GaP$_{0.67}$Sb$_{0.33}$ (a), GaP$_{0.40}$Sb$_{0.60}$ (b), GaSb (c) samples before CMP.

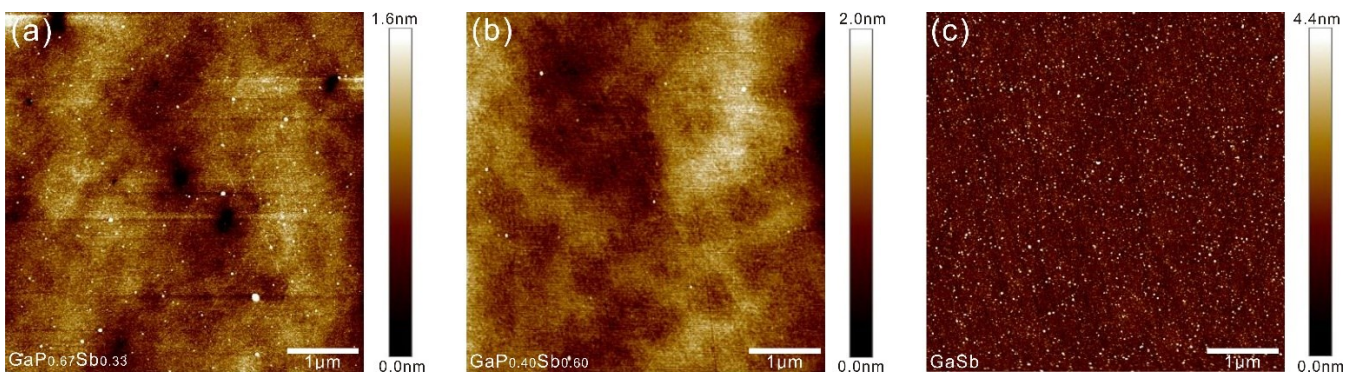

Fig. S4. Atomic force microscopy (AFM) images of the GaP$_{0.67}$Sb$_{0.33}$ (a), GaP$_{0.40}$Sb$_{0.60}$ (b), GaSb (c) samples after CMP.

**Ellipsometry Measurement**

The optical constants of the $GaP_{0.67}Sb_{0.33}$, $GaP_{0.40}Sb_{0.60}$, GaSb samples were measured by variable angle spectroscopic ellipsometry (VASE) at room temperature in the 0.58–5 eV photon energy region. The angles of incidence were set to 60° and 70°. A Tauc-Lorentz model with 2 oscillators was used to fit the ellipsometry data and extract the absorption coefficient value. Fig.S5 show the fitting results for the $GaP_{0.67}Sb_{0.33}$, $GaP_{0.40}Sb_{0.60}$, GaSb samples, respectively. The red and blue lines correspond to experimental spectra where Is and Ic parameters are represented. Is and Ic are related to the well-known ellipsometry variables ψ (amplitude component) and Δ (phase difference) through the following relations: $I_s = \sin(2\psi)\cdot\sin(\Delta)$, $I_c = \sin(2\psi)\cdot\cos(\Delta)$. The black lines correspond to the theoretical curves after adjusting the parameters of the Tauc-Lorentz model. From this model, the refractive index (n), extinction coefficient (k) were extracted, as shown in Fig.S6. In order to further integrate and verify the experimental data, ellipsometry measurement (with incidence angle 70°) was taken on one GaP/Si sample and the corresponding optical constants extracted based on the Tauc-Lorentz model are shown in Fig.S7. The deduced absorption curves of GaP and GaSb based on ellipsometry measurements were compared by the experimental and theoretical data in the reference [1] (Fig.S8), which show good compatibilities.

The absorption spectra were employed to plot the Tauc's curve ($(\alpha h\nu)^k$ vs $h\nu$) for the four GaPSb samples (Figure S9) based on the Tauc's law:

$$(\alpha h\nu)^k = C(h\nu - E_g)$$

where: k=1/2 for indirect bandgap GaP and k=2 for direct bandgap $GaP_{0.67}Sb_{0.33}$, $GaP_{0.40}Sb_{0.60}$, GaSb which are verified by the tight binding calculation in the following part, α the absorption coefficient, h the Planck constant, ν the photon frequency and C a constant. Absorption coefficient range 3000-10000 $cm^{-1}$ were used for bandgap evaluation of the direct bandgap $GaP_{0.67}Sb_{0.33}$, $GaP_{0.40}Sb_{0.60}$, GaSb samples. While for GaP with indirect bandgap, the absorption coefficient corresponding to the bandgap is smaller and the absorption coefficient range was chosen at 100-5000 $cm^{-1}$. The optical band gap can be obtained by linear extrapolation of the straight-line portion to α=0. Finally, the band gap ranges of the four samples were obtained: $E_g$=2.25±0.04 eV for GaP; $E_g$=1.70±0.06 eV for $GaP_{0.67}Sb_{0.33}$; $E_g$=1.04±0.08 eV for $GaP_{0.40}Sb_{0.60}$, and $E_g$=0.68±0.09 eV for GaSb (as shown in the Fig.S9), which are consistent with bandgaps reported for the metamorphic growth of GaPSb on InP substrate[2].

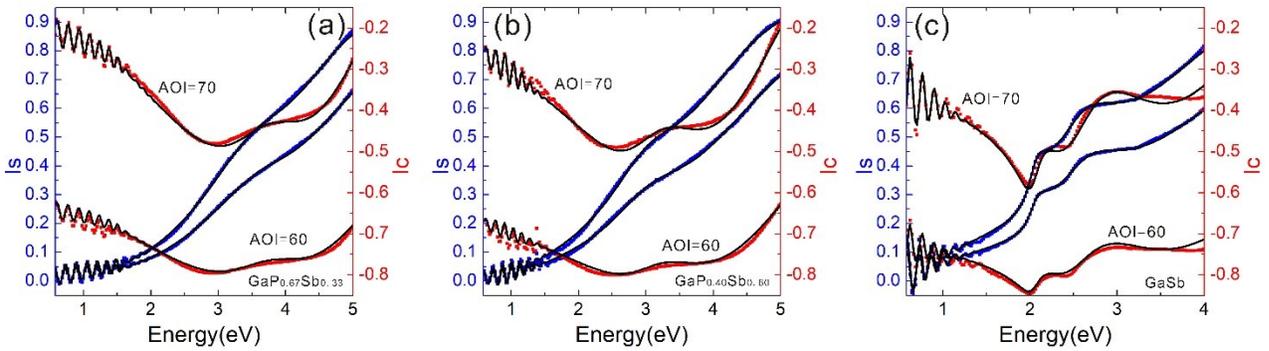

Fig. S5. Experimental ellipsometry spectra of Is and Ic two incidence angles (red and blue lines) and comparison with theoretical curves by using a 2-oscillators Tauc-Lorentz model (black lines) for $GaP_{0.67}Sb_{0.33}$ (a), $GaP_{0.40}Sb_{0.60}$ (b), GaSb (c) three samples, respectively.

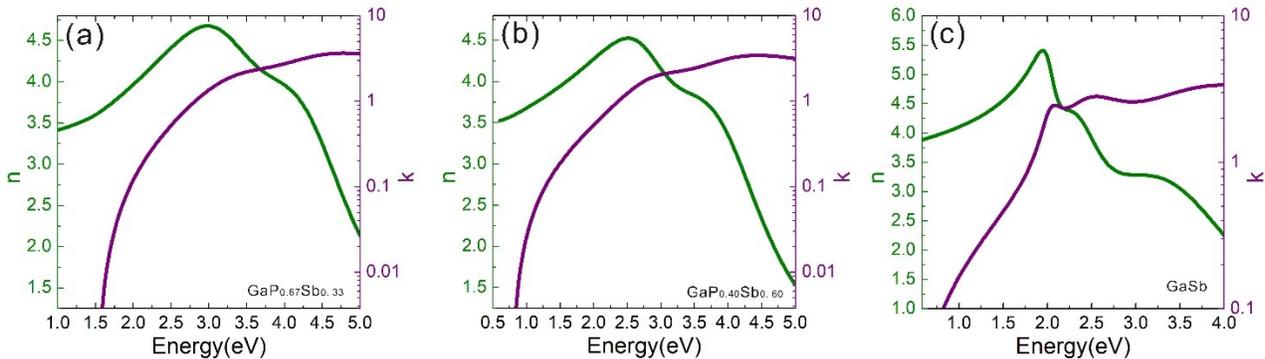

Fig. S6. n (optical index real part) and k (optical index imaginary part) optical constants of GaP$_{0.67}$Sb$_{0.33}$ (a), GaP$_{0.40}$Sb$_{0.60}$ (b), GaSb (c) extracted from the fitting of Fig.S3 (a),(b),(c), respectively.

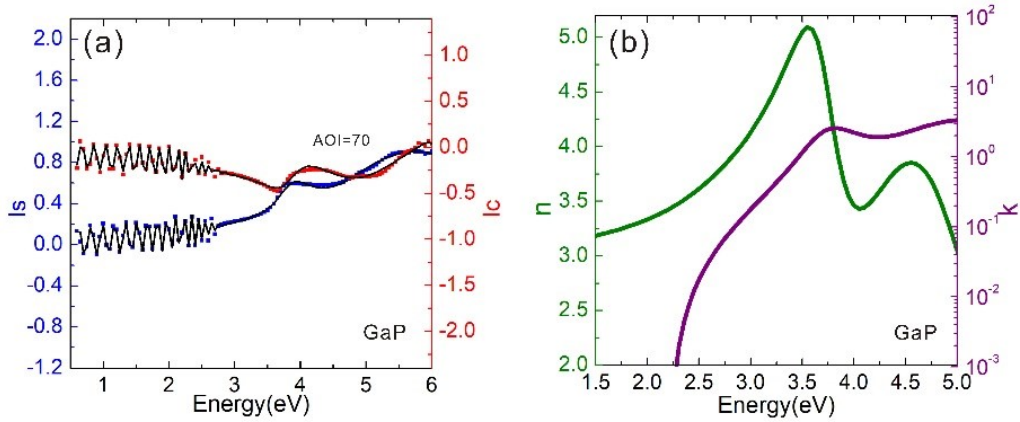

Fig. S7. (a) Experimental ellipsometry spectra of Is and Ic two incidence angles (red and blue lines) and comparison with theoretical curves by Tauc-Lorentz model (black lines) for GaP. (b) n (optical index real part) and k (optical index imaginary part) optical constants of GaP extracted from the fitting of Fig.S7 (a).

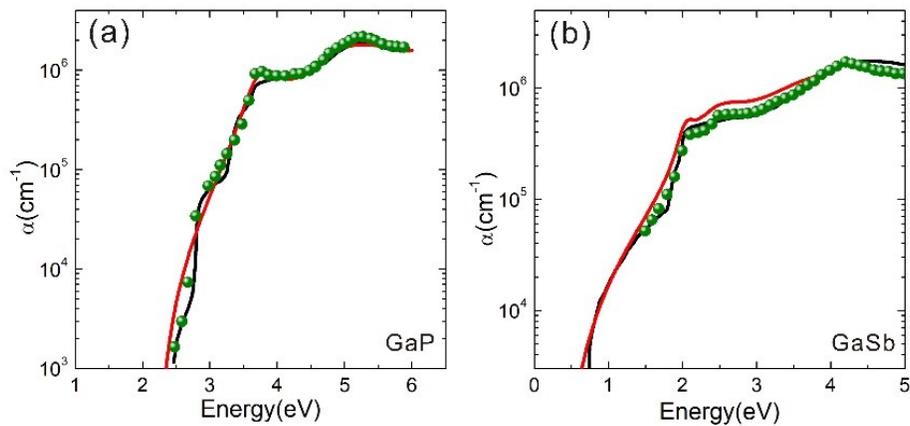

Fig. S8. Optical absorption spectra of GaP (a) and GaSb (b). The red lines were deduced from ellipsometry measurements based on our samples. The green experimental data dots and the black theoretical lines are obtained from the literature [1].

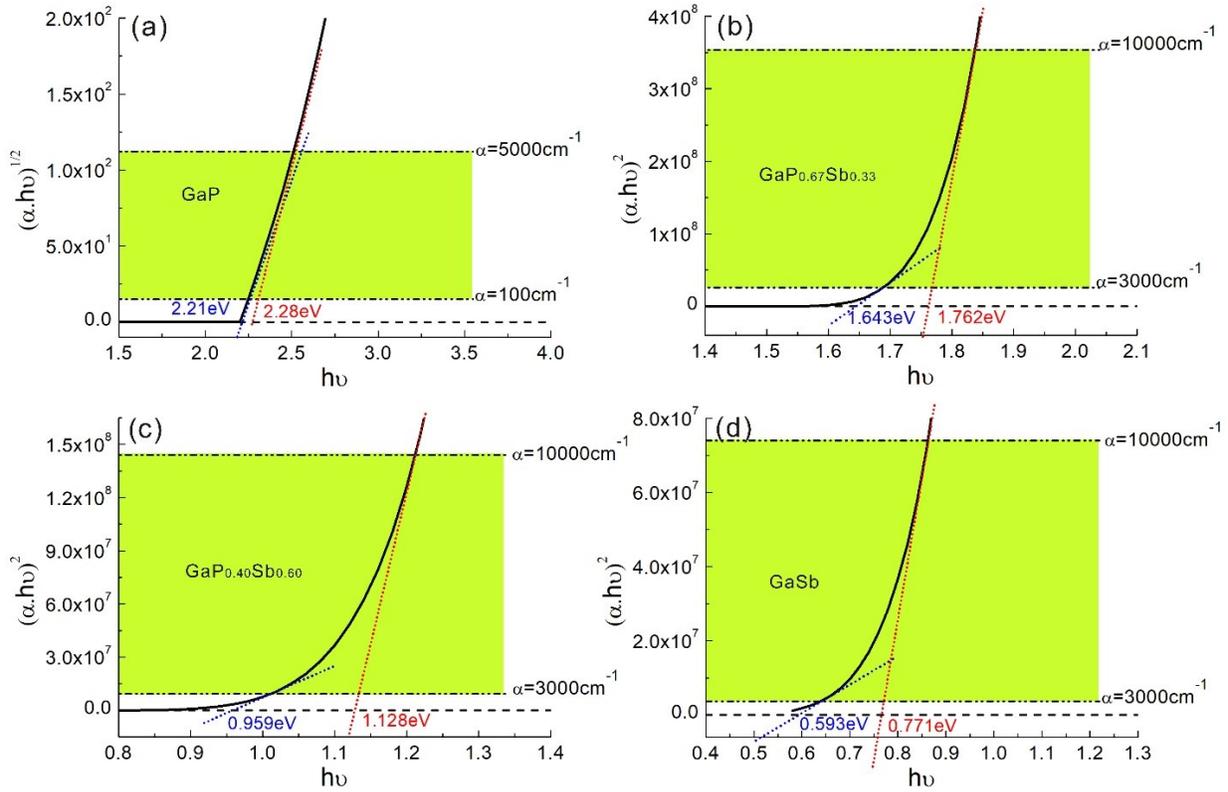

Fig. S9. Tauc plot of $(\alpha h\nu)^k$ versus photon energy ($h\nu$) for GaP (a), $GaP_{0.67}Sb_{0.33}$ (b), $GaP_{0.40}Sb_{0.60}$ (c), GaSb (d) samples.

**Tight Binding Calculation**

The band diagram of $GaP_{1-x}Sb_x$ alloys were calculated using an extended basis sp3d5s* tight binding Hamiltonian [3]. This method was proved to provide a band structure description with a sub-millielectronvolt precision throughout the Brillouin zone of binary cubic III-V and II-VI [4] semiconductors including quantum heterostructures[5] and surfaces[6]. From the tight binding parameters of GaP and GaSb binary compounds [3], a virtual crystal approximation is performed to obtain band structure of $GaP_{1-x}Sb_x$ alloy at different Sb content. The tight binding parameters of the virtual crystal are an arithmetic mean of the constituent materials weighted to their concentration [7] except for the diagonal matrix elements related to the atomic energies of anion "s-type" and 'p-type" states. For these states, bowing parameters are introduced to model the strong bowing of Γ bandgap of 2.7 eV[7]. With a bowing parameter equal to 9.0 eV for s-state and equal to 2.8 eV for p-state, the experimental bandgaps and absorption curves for different $GaP_{1-x}Sb_x$ alloys are nicely reproduced.

The room temperature band structure of GaP, $GaP_{0.67}Sb_{0.33}$, $GaP_{0.40}Sb_{0.60}$, GaSb semiconductors as representatives are shown in the Fig.S10, which has been put temperature effect of the energy shift [8] into consideration. We can see the $GaP_{0.67}Sb_{0.33}$, $GaP_{0.40}Sb_{0.60}$, GaSb semiconductors are direct band transitions and the bandgap energies are 1.673eV, 1.064eV, 0.726eV, respectively, and GaP is indirect band transitions and the bandgap energies are 2.268eV, which are in good agreement with the ellipsometry results.

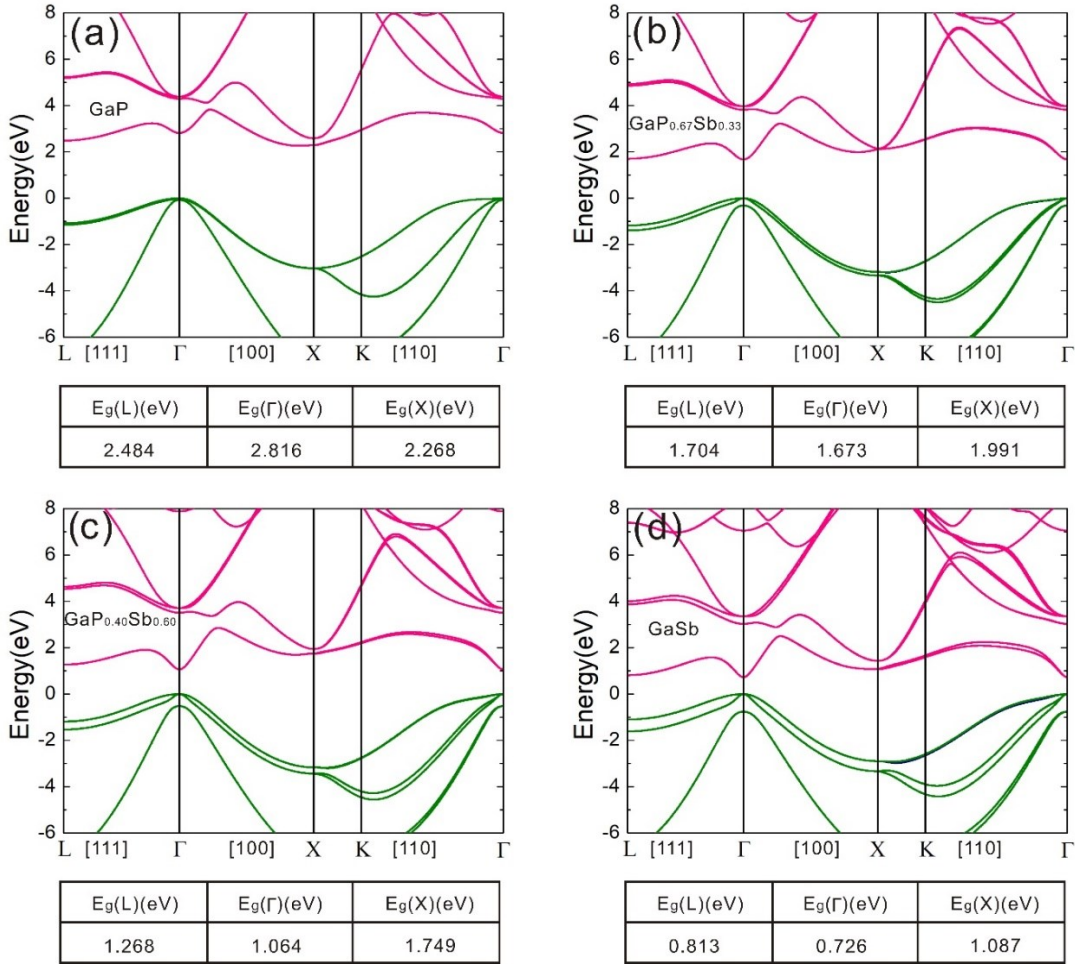

Fig. S10. Band structures of the bulk unstrained GaP (a), $GaP_{0.67}Sb_{0.33}$ (b), $GaP_{0.40}Sb_{0.60}$ (c), GaSb (d) obtained by TB calculation at room temperature.

**Reference**

[1] S. Adachi, Optical dispersion relations for GaP, GaAs, GaSb, InP, InAs, InSb, AlxGa1−xAs, and In1−xGaxAsyP1−y, J. Appl. Phys. 66 (1989) 6030–6040. doi:10.1063/1.343580.
[2] S. Loualiche, A. Le Corre, S. Salaun, J. Caulet, B. Lambert, M. Gauneau, D. Lecrosnier, B. Deveaud, GaPSb: A new ternary material for Schottky diode fabrication on InP, Appl. Phys. Lett. 59 (1991) 423–424.
[3] J.-M. Jancu, R. Scholz, F. Beltram, F. Bassani, Empirical spds* tight-binding calculation for cubic semiconductors: General method and material parameters, Phys. Rev. B. 57 (1998) 6493.
[4] S. Boyer-Richard, C. Robert, L. Gérard, J.-P. Richters, R. André, J. Bleuse, H. Mariette, J. Even, J.-M. Jancu, Atomistic simulations of the optical absorption of type-II CdSe/ZnTe superlattices, Nanoscale Res. Lett. 7 (2012) 543.
[5] R. Scholz, J.-M. Jancu, F. Beltram, F. Bassani, Calculation of Electronic States in Semiconductor Heterostructures with an Empirical spds* Tight-Binding Model, Phys. Status Solidi B. 217 (2000) 449–460.
[6] F. Sacconi, A. Di Carlo, P. Lugli, M. Stadele, J.-M. Jancu, Full band approach to tunneling in MOS structures, IEEE Trans. Electron Devices. 51 (2004) 741–748.
[7] F. Raouafi, R. Samti, R. Benchamekh, R. Heyd, S. Boyer-Richard, P. Voisin, J.-M. Jancu, Optical properties of potential-inserted quantum wells in the near infrared and Terahertz ranges, Solid State Commun. 236 (2016) 7–11.
[8] Y.P. Varshni, Temperature dependence of the energy gap in semiconductors, Physica. 34 (1967) 149–154.